\documentclass[aps,prb,preprint,superscriptaddress]{revtex4}

\usepackage{graphicx}
\usepackage{amsmath}
\usepackage{subfigure}
\usepackage[german,english]{babel}
\begin{document}

\title{Melting of graphene: from two to one dimension}

\author{K V Zakharchenko, Annalisa Fasolino, J H Los, M I Katsnelson}
\affiliation{Radboud University of Nijmegen, Institute for
Molecules and Materials, Heijendaalseweg 135, 6525 AJ Nijmegen,
The Netherlands}



\date{\today}

\begin{abstract}
The high temperature behaviour of graphene is studied by atomistic simulations based on an accurate interatomic potential for carbon. We find that clustering of Stone-Wales defects and formation of octagons are the first steps in the process of melting which proceeds via the formation of carbon chains. The molten state forms a three-dimensional network of entangled chains  rather than a simple liquid. The melting temperature estimated from the two-dimensional Lindemann criterion and from extrapolation of our simulation for different heating rates is about 
$4900$~K.
\end{abstract}
\maketitle

The structural properties of graphene are as exceptional as its electronic structure \cite{1}.  Being the simplest  two-dimensional (2D) membrane, graphene is a natural benchmark of our understanding of physics in 2D \cite{2,3}. Melting in 2D is usually described in terms of creation of topological defects, like unbound disclinations and dislocations \cite{4}. In the hexagonal lattice of graphene, typical disclinations are pentagons (5) and heptagons (7) while dislocations are 5-7 pairs. In carbon systems, pairs of 5-7 dislocations that may be created by rotation of a carbon bond are called Stone-Wales (SW) defects. SW defects are non-topological defects since they have Burger vector and Frank vector equal to zero\cite{Hirth} and thus do not introduce any torsion or curvature in the system. Although graphene is a 2D crystal, it bears also many analogies to other graphitic structures such as fullerenes and nanotubes for which the melting has been studied in Refs.\onlinecite{Kim} and \onlinecite{Stocks,Kowaki} respectively. Since it is known that melting of small particles is essentially different from that of bulk crystals, the relation between the melting mechanisms in fullerenes and nanotubes and that in graphene  is not a priory known. Besides, fullerenes possess an intrinsic curvature that might drastically affect the whole picture of melting\cite{Nelson1983}.  
Our atomistic simulations give a scenario of the melting of graphene as the decomposition of the 2D crystal in a 3D network of 1D chains. The molten phase is similar to the one found for fullerenes\cite{Kim}. Similar structures can be seen in simulations of nanotubes\cite{Stocks,Kowaki}. 
We find that, for graphene, a crucial role is played by SW defects. It is the clustering of SW defects that triggers the formation of octagons which are the precursors for the spontaneous melting around $T_m \approx 4900$ K in our simulations.  This temperature is higher than the value of 4000 K found for fullerenes\cite{Kim} and close to the 4800 K found for nanotubes\cite{Stocks}, but not as high as the value 5800 K estimated for graphene by extrapolation to infinite radius nanotubes in Ref.\onlinecite{Kowaki}. 

We study the melting of graphene by Monte Carlo simulations in the $NVT$ and $NPT$ ensembles (constant number of particles $N$, constant volume $V$ or constant pressure $P$ and constant temperature $T$)  with periodic boundary conditions in the plane for samples of $N=1008$,  $N=4032$ and $N=16128$ atoms. The interatomic interactions are calculated with the LCBOPII potential \cite{5} that we have shown to describe well the elastic and thermal properties of graphene \cite{6} as well as the liquid phase \cite{5}. It is important to note that this potential allows bond breaking and formation  with realistic energy barriers. Also it gives a much better description of the lattice dynamics\cite{lendertjan} than the Tersoff potential used in Ref.\onlinecite{Stocks}.  To speed up equilibration and improve the sampling of long wavelength out-of-plane distortions, we also used collective “wave moves” as described in Ref.~\onlinecite{7}.  More details on the computational procedure can be found in our previous works \cite{3,6,7}. To estimate the melting temperature $T_m$ we have performed simulations where the temperature is increased at slower and slower heating rates as we will describe. 

\begin{figure}[t]
\begin{center}
\includegraphics[width=0.8\linewidth]{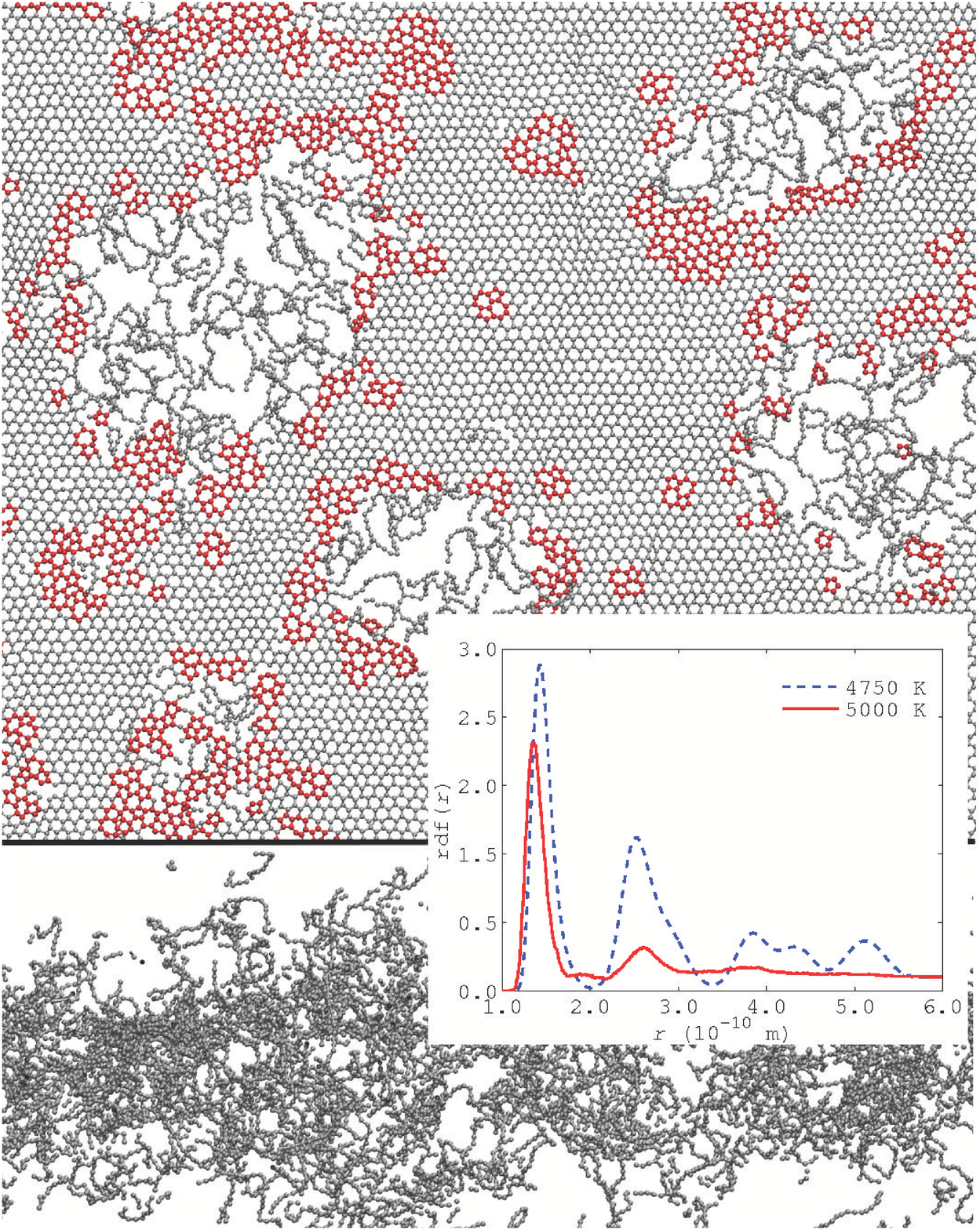}
\end{center}
\caption{\label{exFig1} Top: Snapshot of graphene at $T=5000$~K during the melting process in a $NPT$ simulation ($N = 1008$, $P=0$). All pentagons and heptagons are marked in red. Bottom: side view of the same sample when the melting is completed. Insert: radial distribution function (rdf) at $T=4750$~K and $T=5000$~K, below and above melting. After melting, the first peak is shifted to smaller distances, reflecting the chain formation and further structure, typical of the crystalline phase, is washed out.}

\end{figure}

In Fig.~\ref{exFig1} we show a typical configuration of graphene on the way to melting at $5000$~K~$> T_m$. The coexistence of crystalline and melted regions indicates a first order phase transition. The most noticeable features are the puddles of graphene that have melted into chains. The melted areas are surrounded by disordered 5-7 clusters, resulting from the clustering and distortion of SW defects. They have the smallest formation energy and start appearing at $\approx$ 3800 K. In Fig.~\ref{exFig1} also isolated and pairs of SW defects are present whereas we never observe isolated pentagons, heptagons or 5-7 dislocations. Possibly,  the  less energetically favourable ring structures observed in the simulations of 
nanotubes~\cite{Stocks} are due to a less accurate description of the energetics given by the Tersoff potential. In comparison with the previous studies for fullerenes and nanotubes, the much larger size of 
our samples allows to investigate the role of the spontaneous formation 
and annihilation of SW defects in the melting process.  In Ref.\onlinecite{Kim}, the effect of SW defects on the melting temperature was studied by adding SW defects by hand. 

\begin{figure}[t]
\begin{center}
\includegraphics[width=0.8\linewidth]{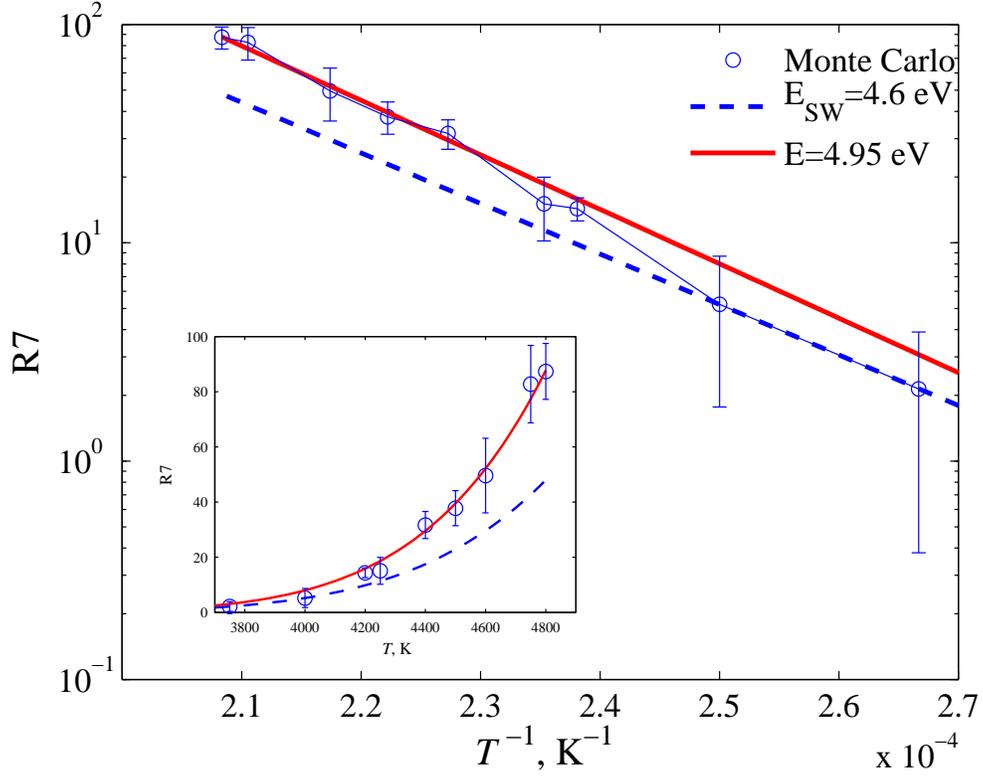}
\end{center}
\caption{\label{newFig2} Average number of heptagonal rings (R7) in the $NVT$ ensemble as a function of temperature for the $N=16128$ sample. The solid line shows the Arrhenius behaviour with the formation energy of SW defects given by LCBOPII, $E_{SW}= 4.6$ eV which gives a good description  only up to $E\approx$ 4000 K whereas a fit to the data for $T >4400$ (dashed line) gives a higher formation energy of $\approx 4.95$ eV, due to SW clusterization. }

\end{figure}

\begin{figure}[t]
\begin{center}
\includegraphics[width=0.7\linewidth]{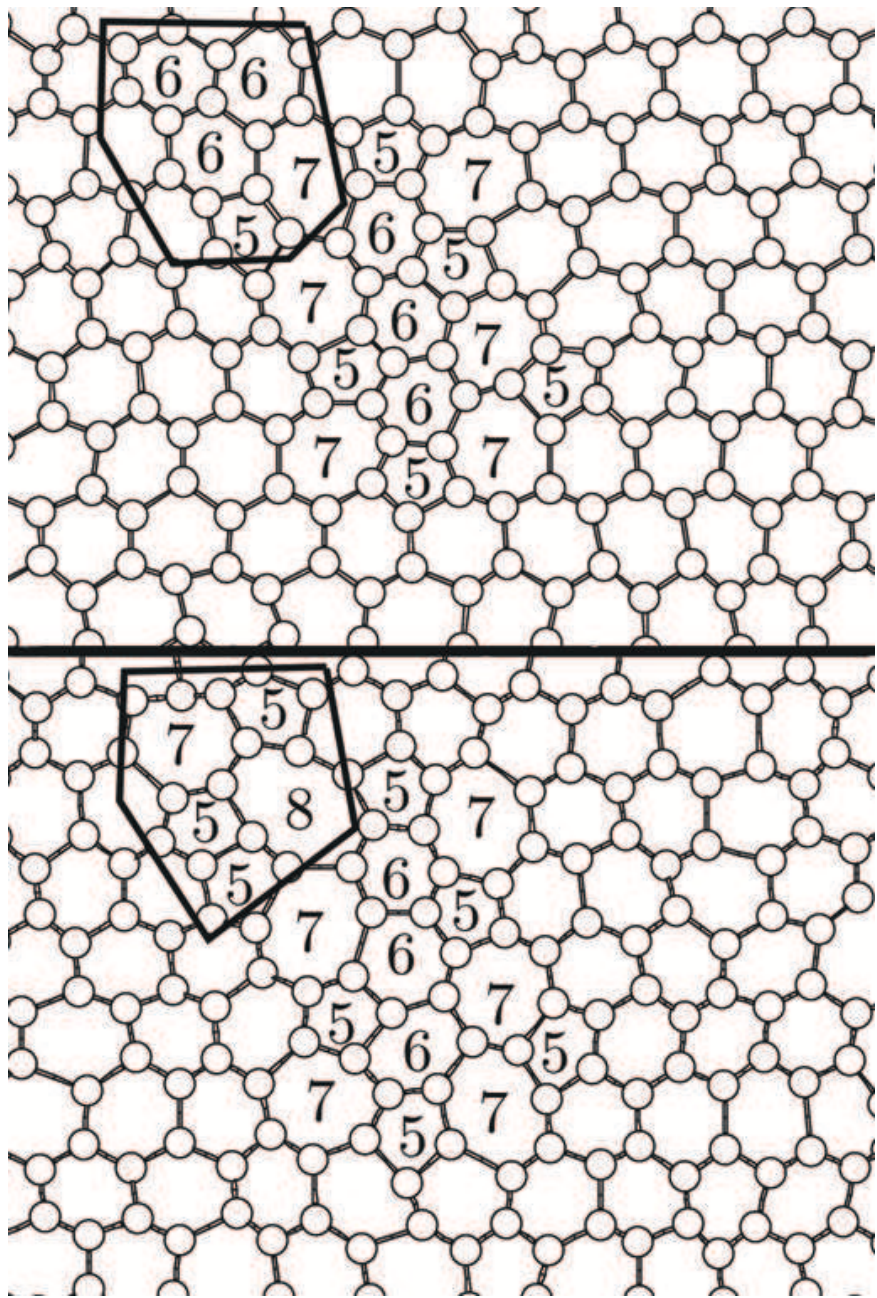}
\end{center}
\caption{\label{exFig2} One typical transformation within a cluster of SW defects (see text). }
\end{figure}

In Fig.~\ref{newFig2} we show the average number of heptagons\cite{note} as a function of temperature together with the Arrhenius behaviour with the formation energy $E_{SW}=4.6$ eV given by LCBOPII\cite{5}. We see that only for  $T < 4000$ K the fit is quite good which demonstrates that all heptagons are part of almost isolated SW defects as also confirmed by visual inspection of typical configurations. At higher temperatures, deviations indicating a higher formation energy per heptagon signals the observed clustering of SW. We find that 5-7 clusters act as nuclei for the melting. This is in contrast with graphite where melting is initiated by interplanar covalent bond formation. Close inspection shows that regions with 5-7 clusters favor the transformation of three hexagons into two pentagons and one octagon (Fig.~\ref{exFig2}) that we never see occurring in the regular hexagonal lattice far from the 5-7 clusters or near isolated SW defects. The octagons, in turn, are the precursors for the formation of larger rings. Due to the weakening of the bonds forming the relatively small angles in the pentagons around them, the atoms forming these larger rings tend to detach from the lattice and form chains. 

When melting is completed the carbon chains form an entangled 3D network with a substantial amount of three-fold coordinated atoms, linking the chains. Therefore, this low density structure reminds rather a polymer gel than a simple liquid. In fact the radial distribution function shown in the inset of Fig.~\ref{exFig1} displays a very sharp peak at distances smaller than the interatomic distance in graphene, due to the formation of shorter double bonds in the chains. In typical simple liquids, instead, the first peak shifts to larger values of interatomic distances when going from a crystal to a liquid \cite{8}. This high temperature phase has also been found for molten fullerenes\cite{Kim}.

\begin{figure}[t]
\begin{center}
\includegraphics[width=0.7\linewidth]{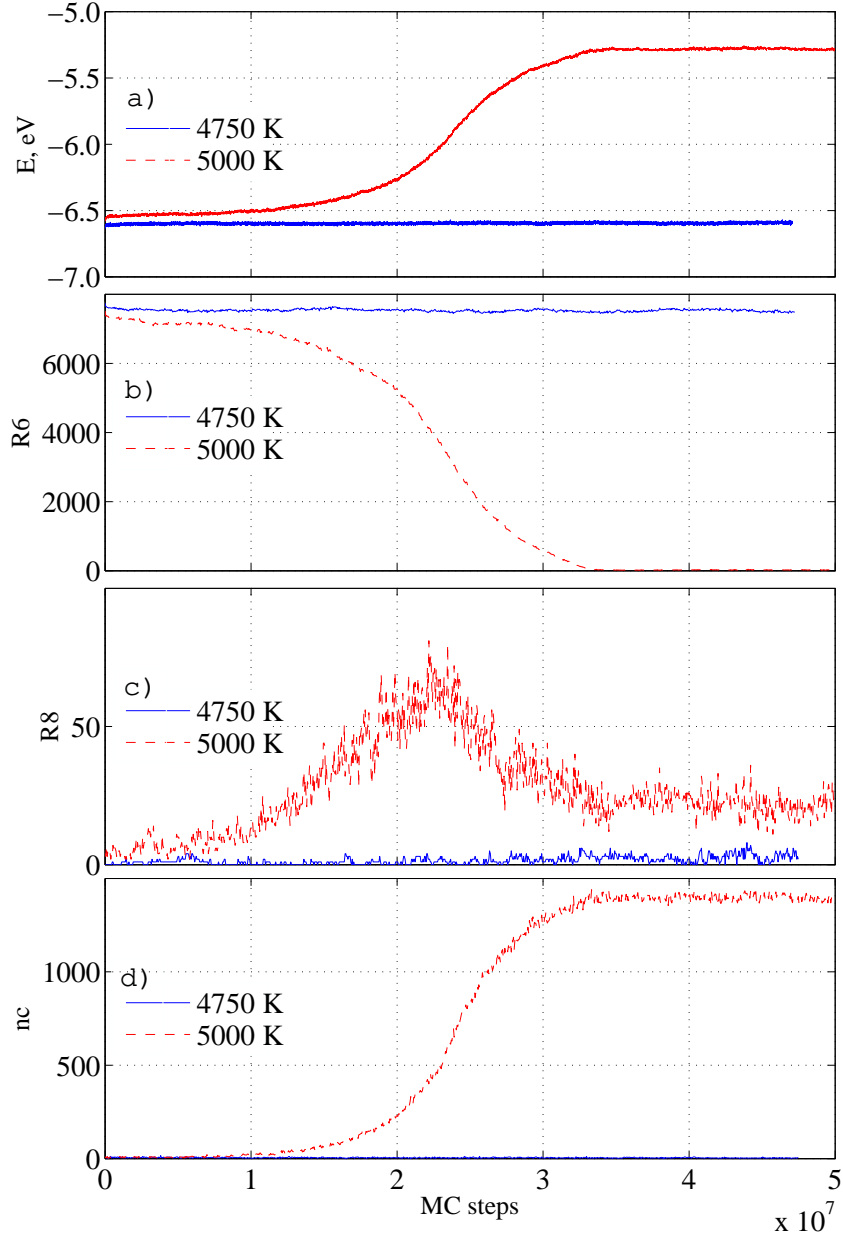}
\end{center}
\caption{\label{exFig3} Behaviour of several quantities as a function of the number of Monte Carlo steps starting from a solid flat graphene layer ($N = 16128$) at $T = 4750$ K (below melting) and at $T = 5000$ K where melting occurs within about $10^7$ Monte Carlo steps around step $2.5\times10^7$. From top to bottom: a) Total potential energy $E$ in eV/atom; b) Number of six-member rings R6; c) Number of 8-member rings R8. Eight-member rings are formed close to clusters of SW defects. R8 increases when melting starts and decreases when chains are formed, illustrating the melting mechanism (see text); d) Number of chains $nc$ with more than $3$ connected two-fold coordinated atoms.}
\end{figure}

In Fig.~\ref{exFig3} we show the evolution of the potential energy and structural properties of an initially flat, graphene layer as a function of Monte Carlo steps (1 step is equal to $N$ displacement trials) at $T=4750$ K (below melting) and at $T=5000$ K where melting occurs within about $10^7$ Monte Carlo steps around step $2.5\times10^7$. We see that the rise of the potential energy at melting, is mirrored by the growth of the number of chains (nc) at the expenses of the six-member rings (R6) of the crystalline phase. The number of eight-member rings (R8) that are formed close to clusters of SW defects, instead,  increases when melting starts and decreases when chains are formed, illustrating the melting mechanisms described previously. 

\begin{figure}[t]
\begin{center}
\includegraphics[width=1.0\linewidth]{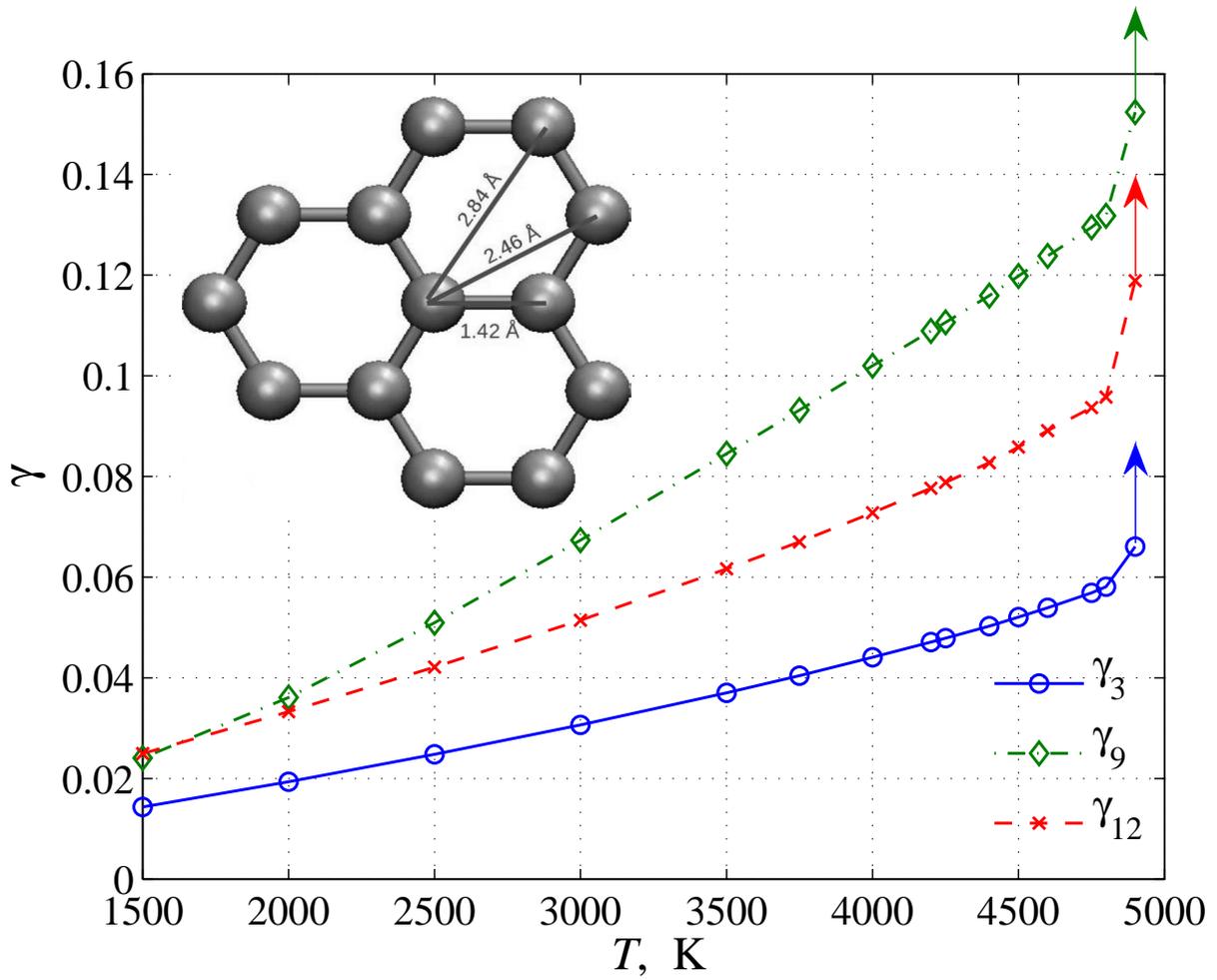}
\end{center}
\caption{\label{exFig4} Temperature dependence of $\gamma_3$,$\gamma_9$ and $\gamma_{12}$ calculated including the closest three, nine and twelve neighbours (see inset). The last points indicated by arrows correspond to the onset of the  liquid phase for which these quantities diverge for infinite systems. }
\end{figure}

The Lindemann criterion \cite{8,9}  $\sqrt{\left < u^2\right >_m}/d\approx 0.2$
(where $\left <u^2 \right >_m$  is the mean-square atomic displacement at the melting temperature and $d$ is the interatomic distance at $T=0$) is commonly used to estimate the melting temperature in 3D systems. Since in 2D the mean square displacement $\left <u^2 \right >$   is divergent at finite temperatures \cite{10}, we need to consider differences of atomic displacements \cite{11,12}. Adapting the melting criterion used in Ref.\onlinecite{11} to the honeycomb lattice of graphene we define the average quantity  
\begin{equation}
\gamma_n=\frac{1}{a^2} \left <\left| {\bf r}_i - \frac{1}{n} \sum_j {\bf r}_j \right|^2 \right>
\end{equation}
where $ a = 1/\sqrt{\pi \rho_0} $ where $ \rho_0 $ is the 2D particle density at $T=0$~K,
$ {\bf r}_i $ is the position of the $i$-th atom and where the sum over $j$ runs over the 
$n$ atoms closest to atom $i$.
In Fig.~\ref{exFig4} we show  $\gamma_n$ for $n = 3$, $9$ and $12$, namely, including one, two and three coordination spheres (see the inset in Fig.~\ref{exFig4}).  Since the difference between second and third neighbor distances is relatively small, the distinction between second and third neighbours becomes  fuzzy at very high temperatures as one can see from the merging of the related peaks in the radial distribution in the inset of Fig.~\ref{exFig1}. Therefore we think that $\gamma_{12}$ is more meaningful than $\gamma_9$. Interestingly, as shown in Fig.~\ref{exFig4}, melting occurs when  $\gamma_{12} \approx 0.1$  as found for the strictly 2D triangular lattice in Ref.~\cite{11}. Instead, $\gamma_3$ remains much smaller due to the rigidity of the covalent nearest-neighbour bonds.  

\begin{figure}[t]
\begin{center}
\includegraphics[width=1.0\linewidth]{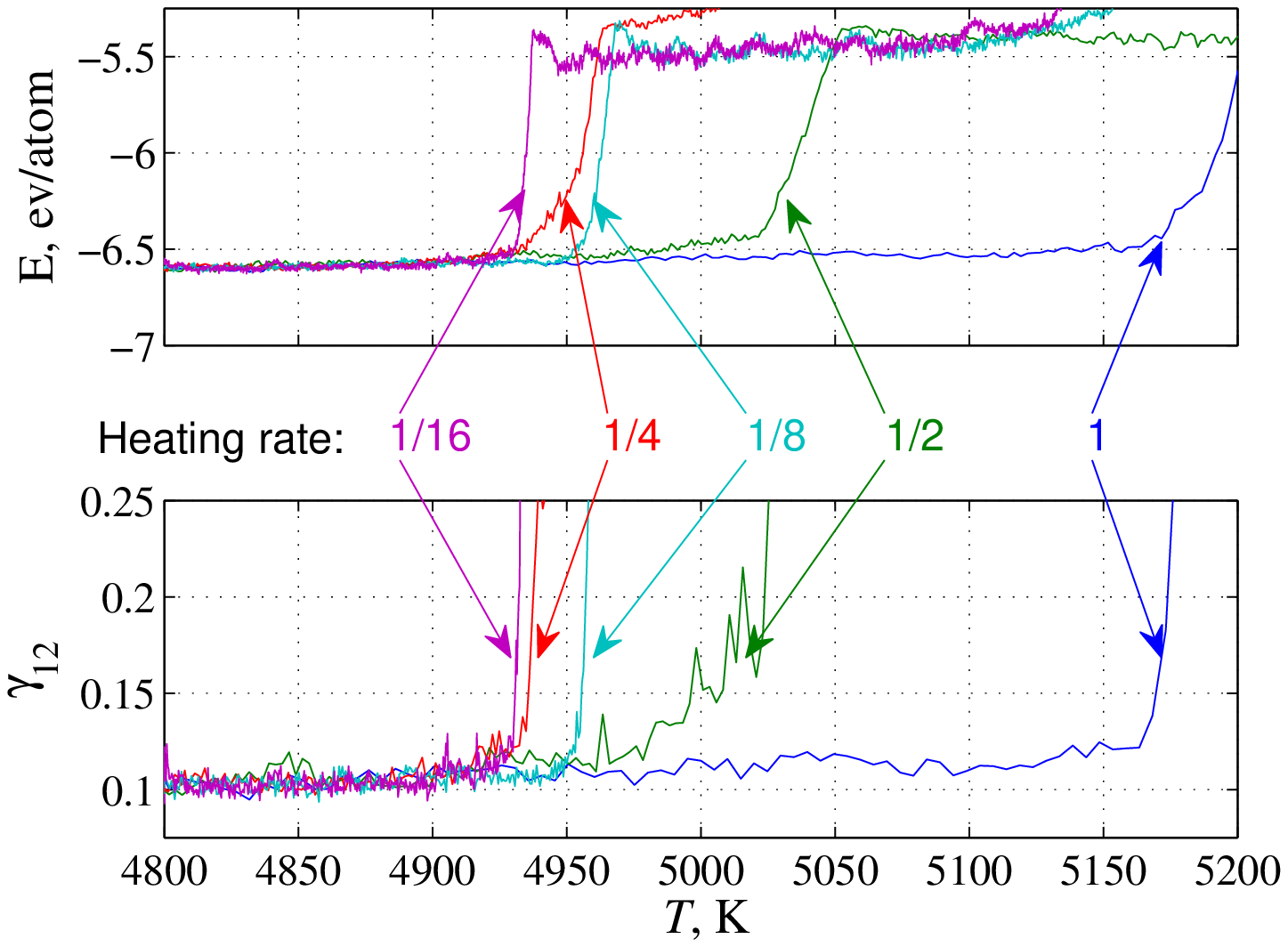}
\end{center}
\caption{ \label{exFig5} Potential energy (a) and  $\gamma_{12}$(b) as a function of temperature for graphene layer ($N = 1008$) for five different heating rates, each two times slower than the previous one, namely $1$, $1/2$, $1/4$, $1/8$, $1/16$  degree Kelvin per $1000$  Monte Carlo steps as indicated by the labels. }
\end{figure}

To estimate the melting temperature we consider the effect of heating the sample at different rates. In Fig.~\ref{exFig5} we show the potential energy per atom (Fig.~\ref{exFig5}a) and  $\gamma_{12}$ (Fig.~\ref{exFig5}b) for several heating rates, each half of the previous one. One can see that the temperature at which the energy and   $\gamma_{12}$ suddenly grow, signaling the melting, moves from about $5200$ K towards the left saturating slightly above $4900$ K. 

The closest system to graphene is graphite. The melting temperature of graphite has been extensively studied experimentally at pressures around $10$ GPa and the results present a large spread between $4000$ K and $5000$ K \cite{13}. With LCBOPII, free energy calculations give $T_m = 4250$ K, almost independent of pressure between $1$ and $20$ GPa \cite{14}. At zero pressure, however,  graphite sublimates before melting at $3000$ K \cite{13}. Monte Carlo simulations with LCBOPII at zero pressure show that, at $3000$ K, graphite sublimates through detachment of the graphene layers \cite{14}. 
The melting of graphene in vacuum that we have studied here can be thought of as the last step in the thermal decomposition of graphite,  the 2D graphene layers melting into a 3D liquid  network of 1D chains. Interestingly, formation of carbon chains has been observed in the melt zone of graphite under laser irradiation\cite{Hu}
Although the temperature $T=4900$~K of spontaneous melting represents an upper limit for $T_m$, our simulations suggest that $T_m$ of graphene at zero pressure is higher than that of graphite.

\newpage

\newpage

\newpage

\newpage

\newpage

\newpage

\acknowledgments
This work is supported by FOM-NWO, the Netherlands. We thank Jaap Kroes for adapting the code polipy~\cite{note}.

\end{document}